\documentclass[twocolumn,pra,showpacs,aps]{revtex4-1}
\usepackage{amssymb}
\usepackage{amsmath}
\usepackage{graphicx}
\usepackage{epsfig}

\begin{document}

\title{Non-Hermitian interferometer: Unidirectional amplification without
distortion}
\author{C. Li, L. Jin and Z. Song}
\email{songtc@nankai.edu.cn}
\affiliation{School of Physics, Nankai University, Tianjin 300071, China}

\begin{abstract}
A non-Hermitian interferometer can realize asymmetric transmission in the
presence of imaginary potential and magnetic flux. Here, we propose a
non-Hermitian dimer with an unequal hopping rate by an interferometer-like
cluster in the framework of tight-binding model. The intriguing features of
this design are the wave-vector independence and unidirectionality of
scattering, which amplifies wavepacket without distortion and absorbs
incoherent wave without reflection. The absorption relates to the system
spectral singularities, the dynamical behaviors of the spectral
singularities are also investigated analytically and numerically.
\end{abstract}

\pacs{03.65.Nk,05.60.Gg,11.30.Er,42.25.Bs}
\maketitle



\section{Introduction}

Nowadays a non-Hermitian quantum mechanics has emerged as a versatile
platform for exploring the difficulties for fabricating functional devices
in Hermitian regime. The main mechanism is based on the existence of
imaginary potential which has been investigated theoretically \cite%
{Bender,JPA2,Ali,Znojil,Jones,OL07,PRL08a,PRL08b,Joglekar10,Joglekar11,YDChong,HJingPRL2014}
and realized in experiment \cite%
{AGuo,CERuter,Wan,Sun,LFeng,BPeng,LChang,LFengScience,HodaeiScience,NC2015}
as an ideal building block of non-Hermitian system. However, it has been
shown a pure complex potential cannot realize asymmetric transmission \cite%
{Muga,AnnPhys,LXQ}, which is the central goal in many works \cite%
{Ahmed,Ali4,Longhi}. On the other hand, another ideal building block
possessing asymmetric transmission is a asymmetric dimer, which has an
unequal hopping strength in the framework of tight-binding model.

\begin{figure}[tbp]
\includegraphics[bb=50 170 540 810, width=0.43\textwidth, clip]{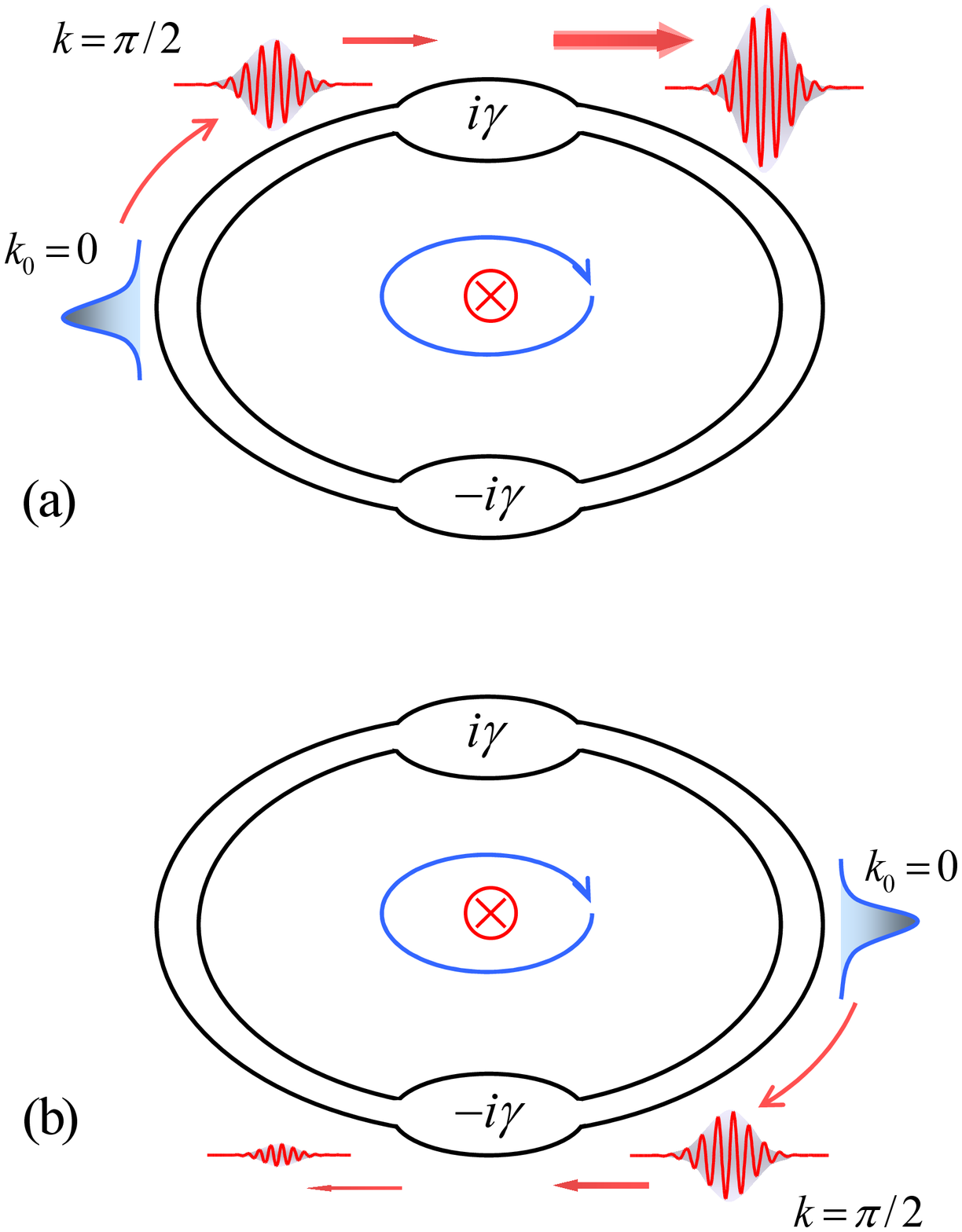}
\caption{(Color online) Schematic illustration of a simple non-Hermitian
configuration to demonstrate the the associated effect of imaginary
potential and magnetic flux on the dynamics of a wavepacket. It a
non-Hermitian ring with the non-Hermiticity arising from the imaginary
potentials. The magnetic flux breaks left-right symmetry. Initially, a
stationary wavepacket (blue) with vector $k_{0}=0$ is located at the left
(a) or right (b) respectively. In both cases, the central momentum of the
wavepacket can be shifted to $\protect\pi /2$\ by a well-prepared flux. It
makes two wavepackets face two different situations. (a) The wavepacket
moves up and encounters the potential $2i$. It acquires an infinite
transmission. (b) The wavepacket moves down and is scattered by the
potential $-2i$ and get a finite transmission.} \label{figure1}
\end{figure}

In this paper, we investigate the possible mechanism of asymmetric
transmission in a non-Hermitian system. We show that a non-Hermitian
interferometer can realize asymmetric transmission due to the combination of
imaginary potential and magnetic flux. As a demonstration, we construct a
non-Hermitian dimer with an unequal hopping rate by an interferometer-like
cluster in the framework of tight-binding model. The intriguing feature of
this design are the wave-vector independent and unidirectional scattering,
which allows the reflectionless amplified transmission of wavepacket without
any distortion. With optimal system parameters, it acts as an absorber for
both coherent and incoherent incident waves. Some dynamical behaviors
related to the spectral singularities are also presented analytically and
numerically.

This paper is organized as follows. In Section \ref{Asymmetric transmission}%
, we depict the physical mechanism of asymmetric transmission arising from
the combination of imaginary potential and magnetic flux. In Section \ref%
{Asymmetric dimer}, we present a interferometer-like scattering center,
which is shown to be equivalent to an asymmetric dimer. In Section \ref%
{Reflectionless amplification} we study the transmission feature of this
design. In Section \ref{Spectral singularity}, the spectral singularity of
the Hamiltonian is examined. Finally, we give a summary in Section \ref%
{Summary}.

\section{Asymmetric transmission}

\label{Asymmetric transmission}

It is well known that a Hermitian scattering center possesses the feature of
symmetric transmission, i.e., the transmission and reflection coefficients
are independent of the input direction of an incident wave. It is true for
the case with a threading magnetic flux, which may break the time reversal
symmetry. On the other hand, it has been shown that, a non-Hermitian
scattering center but with the time reversal symmetry still has symmetric
transmission \cite{LXQ}. A typical scattering center of such kind is a
Hermitian system with additional imaginary potentials. It has been found
that the magnetic flux in a non-Hermitian scattering center may lead to
asymmetric transmission. The mechanism of this behavior can be understood by
the following simple example.

\begin{figure}[tbp]
\includegraphics[bb=22 348 501 802, width=0.43\textwidth, clip]{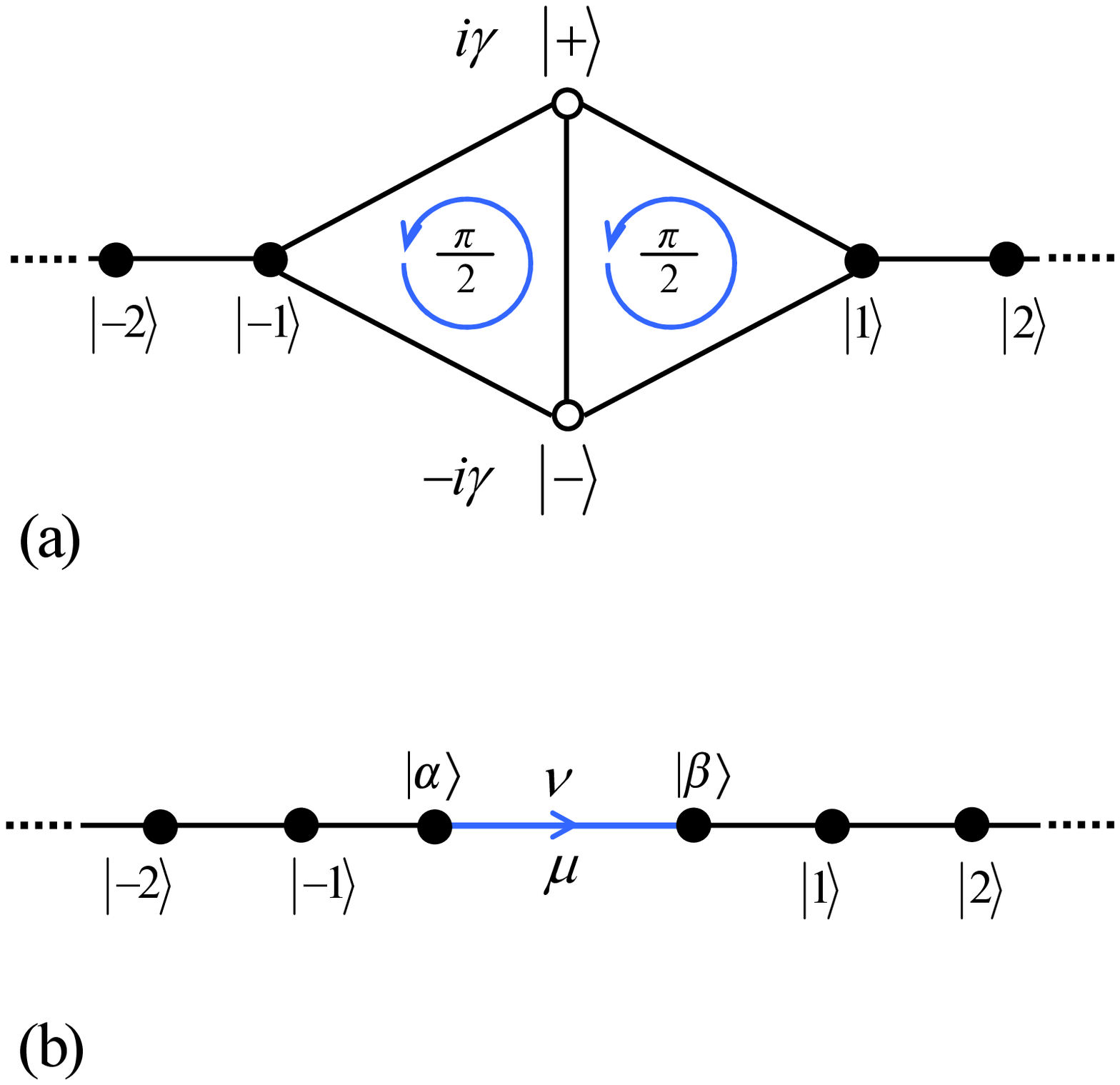}
\caption{(Color online) Sketch of the system considered in this work and
schematic illustration of the equivalent Hamiltonian with asymmetric
transmission. The black bonds denote Hermitian hopping, while the blue
arrowed bond denotes the hopping with asymmetric amplitudes. (a) The
non-Hermitian scattering center embedded in a discrete waveguide. It
consists of Hermitian hopping, magnetic flux, and imaginary potentials. The
coexistence of flux and imaginary potentials breaks the symmetry of
transmission. (b) The equivalent system of (a), in which the scattering
center is simply an asymmetric dimer with unequal hopping strength $(\protect\mu ,\protect\nu )$.}
\label{figure2}
\end{figure}

We start with a simplest non-Hermitian scattering center, an on-site
imaginary potential embedded in the center of an infinite chain with the
Hamiltonian%
\begin{equation}
H_{\mathrm{\gamma }}=H_{\mathrm{lead}}+H_{\mathrm{c}}
\end{equation}%
where%
\begin{equation}
H_{\mathrm{lead}}=-\sum_{j=1}^{\infty }\left( \left\vert j\right\rangle
\left\langle j+1\right\vert +\left\vert -j\right\rangle \left\langle
-j-1\right\vert +\mathrm{H.c.}\right) ,
\end{equation}%
is the Hamiltonian of two leads and the scattering center Hamiltonian%
\begin{equation}
H_{\mathrm{c}}=-\left( \left\vert -1\right\rangle +\left\vert 1\right\rangle
\right) \left\langle 0\right\vert +\mathrm{H.c.}+i\gamma \left\vert
0\right\rangle \left\langle 0\right\vert .
\end{equation}%
The Bethe Ansatz solution \cite{Kim PRB} gives the transmission and
reflection coefficients

\begin{equation}
T_{k}(\gamma )=\frac{4\sin ^{2}k}{\left( 2\sin k-\gamma \right) ^{2}},\text{
}R_{k}(\gamma )=\frac{\gamma ^{2}}{\left( 2\sin k-\gamma \right) ^{2}},
\end{equation}%
for the incident wave from left or right side. We note that $T_{k}$\ is $%
\gamma $- and $k$-dependent but independent of incident direction. Two
typical cases are of $\gamma =-2$\ and$\ 2$ for $k=\pi /2$ , with two
different features, i.e.,
\begin{equation}
T_{\pi /2}(2)=\infty ,\text{ }T_{\pi /2}(-2)=1/4.
\end{equation}%
\ On the other hand, as a Hermitian quantity, the magnetic flux has two
features: i) it can shift the wave vector of a plane wave; ii) it usually
breaks the symmetry in real space. Combining the effects of imaginary
potential and flux, the asymmetric transmission can be realized in
principle. To demonstrate this point, we consider a concrete system, a
non-Hermitian ring, which is schematically illustrated in Fig. \ref{figure1}%
. The non-Hermiticity of the system arises from the imaginary potentials.
The magnetic flux breaks left-right symmetry. In the following we examine
the effects of potential and flux on the dynamics of a wavepacket. Consider
a stationary wavepacket with vector $k_{0}=0$, which is initially located at
the left or right, respectively. In both cases, the central momentum of the
wavepacket can be shifted to $\pi /2$\ by a well-prepared flux \cite{YangS}.
It makes two wavepackets face two different situations. For the left one,\
it is scattered by the potential $2i$\ and get an infinite transmission,
while the right one is scattered by the potential $-2i$ and get a finite
transmission. Apparently, it is due to the left-right\ symmetry breaking
induced by the flux. The imaginary potential takes an important role. If we
replace $i\gamma $ by real number $V$, we have the corresponding
transmission coefficient $T_{k}(V)=4\sin ^{2}k/\left( 4\sin
^{2}k+V^{2}\right) $, which is independent of the sign of $V$ (see Appendix
a). It accords with the exact result for a Hermitian scattering center with
flux \cite{LXQ}. On the other hand, imaginary potentials cannot leads to
asymmetric transmission solely, since the system is left-right symmetry in
the absence of flux.

So far we have given a semi-quantitative analysis about the realization of
asymmetric transmission. For a tight-binding lattice network, the
equivalence between the imaginary potential and the input (output) lead is
proposed \cite{L. Jin10,JL,ZG PRA}. In the following section, we will
construct a simple but efficient system, which will be shown to be
equivalent to an asymmetric dimer, to demonstrate the novel feature of
non-Hermitian system.

\section{Asymmetric dimer}

\label{Asymmetric dimer}

In this section, we will present a concrete scattering center, which is
exactly solvable and deliberately constructed to exhibit unambiguous
asymmetric transmission. This study is of significant not only for the
non-Hermitian quantum mechanics but also for applications in quantum
technology. Now the model we study becomes%
\begin{equation}
H=H_{\mathrm{lead}}+H_{\mathrm{int}}.
\end{equation}%
The concerned scattering center is a non-Hermitian Aharonov-Bohm
interferometer, with the\ Hamiltonian
\begin{eqnarray}
H_{\mathrm{int}} &=&-\frac{1}{\sqrt{2}}\sum_{\sigma =\pm }\left( e^{-i\sigma
\phi }\left\vert -1\right\rangle +e^{i\sigma \phi }\left\vert 1\right\rangle
\right) \left\langle \sigma \right\vert  \notag \\
&&+\delta \left\vert +\right\rangle \left\langle -\right\vert +\mathrm{H.c.}%
+i\gamma \sum_{\sigma =\pm }\sigma \left\vert \sigma \right\rangle
\left\langle \sigma \right\vert ,  \label{original}
\end{eqnarray}%
where parameters $\delta $ and $\gamma $\ are real numbers and $\phi =\pi /4$%
. It is constructed by a Hermitian cluster with additional imaginary
potentials. The geometry of the cluster and the process of simplification
are illustrated in Fig. \ref{figure2}. Recently, the transmission problem
for similar non-Hermitian Aharonov-Bohm 4-site rings is studied in \cite{ZG
PRA,Zeng}.

A tight-binding network is constructed topologically by the sites and
various connections between them. There are three types of basic
non-Hermitian clusters leading to the non-Hermiticity of a discrete
non-Hermitian\ system: i) complex on-site potential denoted as $e^{i\varphi
}\left\vert l\right\rangle \left\langle l\right\vert $; ii) non-Hermitian
dimer denoted as $e^{i\varphi }$($\left\vert l\right\rangle \left\langle
j\right\vert +\mathrm{H.c.}$), where $\varphi $\ is real; iii) asymmetric
hopping amplitude dimer denoted as $\mu \left\vert l\right\rangle
\left\langle j\right\vert +\nu \left\vert j\right\rangle \left\langle
l\right\vert $ with asymmetric parameters $\mu \neq \nu $ being real
numbers, which has been used in modeling a delocalization phenomenon
relevant for the vortex pinning in superconductors \cite{Hatano}. The former
two types of non-Hermitian clusters violate $\mathcal{T}$ symmetry, while
the last one does not. The non-Hermiticity of the present system only arises
from the imaginary potentials. In previous work, it has been shown by an
example that, the first and second types of clusters are transformable with
each other \cite{ZG PRA}.\ In the following, we will show that our model can
be reduced to the third types of dimers by a simplification process.

By taking the linear transformation%
\begin{equation}
\left\{
\begin{array}{c}
\left\vert \alpha \right\rangle =\frac{1}{\sqrt{2}}\left( e^{i\pi
/4}\left\vert +\right\rangle +e^{-i\pi /4}\left\vert -\right\rangle \right) ,
\\
\left\vert \beta \right\rangle =\frac{-i}{\sqrt{2}}\left( e^{i\pi
/4}\left\vert +\right\rangle -e^{-i\pi /4}\left\vert -\right\rangle \right) ,%
\end{array}%
\right.
\end{equation}%
the Hamiltonian of the scattering center is reduced to%
\begin{eqnarray}
H_{\mathrm{int}} &=&-\left( \left\vert -1\right\rangle \left\langle \alpha
\right\vert +\left\vert 1\right\rangle \left\langle \beta \right\vert
\right) +\mathrm{H.c}.  \notag \\
&&+\left( \delta +\gamma \right) \left\vert \alpha \right\rangle
\left\langle \beta \right\vert +\left( \delta -\gamma \right) \left\vert
\beta \right\rangle \left\langle \alpha \right\vert ,
\end{eqnarray}%
which is schematically illustrated in Fig. \ref{figure2}(b). This let us can
rewrite the whole Hamiltonian as an equivalent one%
\begin{eqnarray}
H_{\mathrm{eq}} &=&-\sum_{j=1}^{\infty }\left( \left\vert j\right\rangle
\left\langle j+1\right\vert +\left\vert -j\right\rangle \left\langle
-j-1\right\vert \right)  \notag \\
&&-\left( \left\vert -1\right\rangle \left\langle \alpha \right\vert
+\left\vert 1\right\rangle \left\langle \beta \right\vert \right) +\mathrm{%
H.c}.  \notag \\
&&-\mu \left\vert \alpha \right\rangle \left\langle \beta \right\vert -\nu
\left\vert \beta \right\rangle \left\langle \alpha \right\vert ,
\label{H_eq}
\end{eqnarray}%
for scattering problem. Here parameters%
\begin{equation}
\mu =-\left( \delta +\gamma \right) \text{, }\nu =-\left( \delta -\gamma
\right)
\end{equation}%
are asymmetric hopping amplitudes, exhibiting the feature of asymmetric
transmission in a simple way. In the following sections, we will explore the
features of the asymmetric dimer, which can be applied to the original
system in Eq. (\ref{original}).
\begin{figure*}[tbp]
\centering
\includegraphics[ bb=39 238 420 558, width=0.4\textwidth, clip]{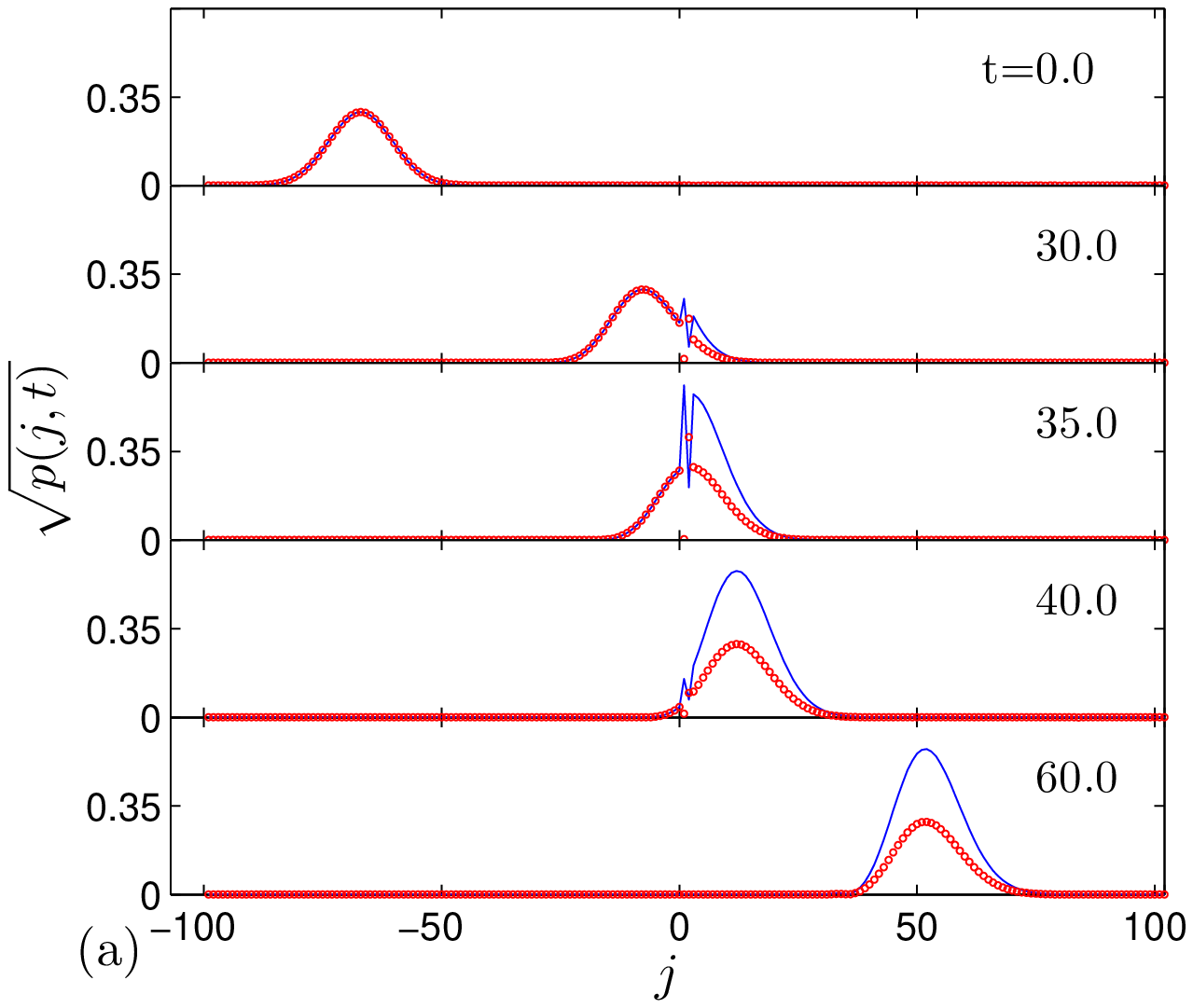} %
\includegraphics[ bb=39 238 420 558, width=0.4\textwidth, clip]{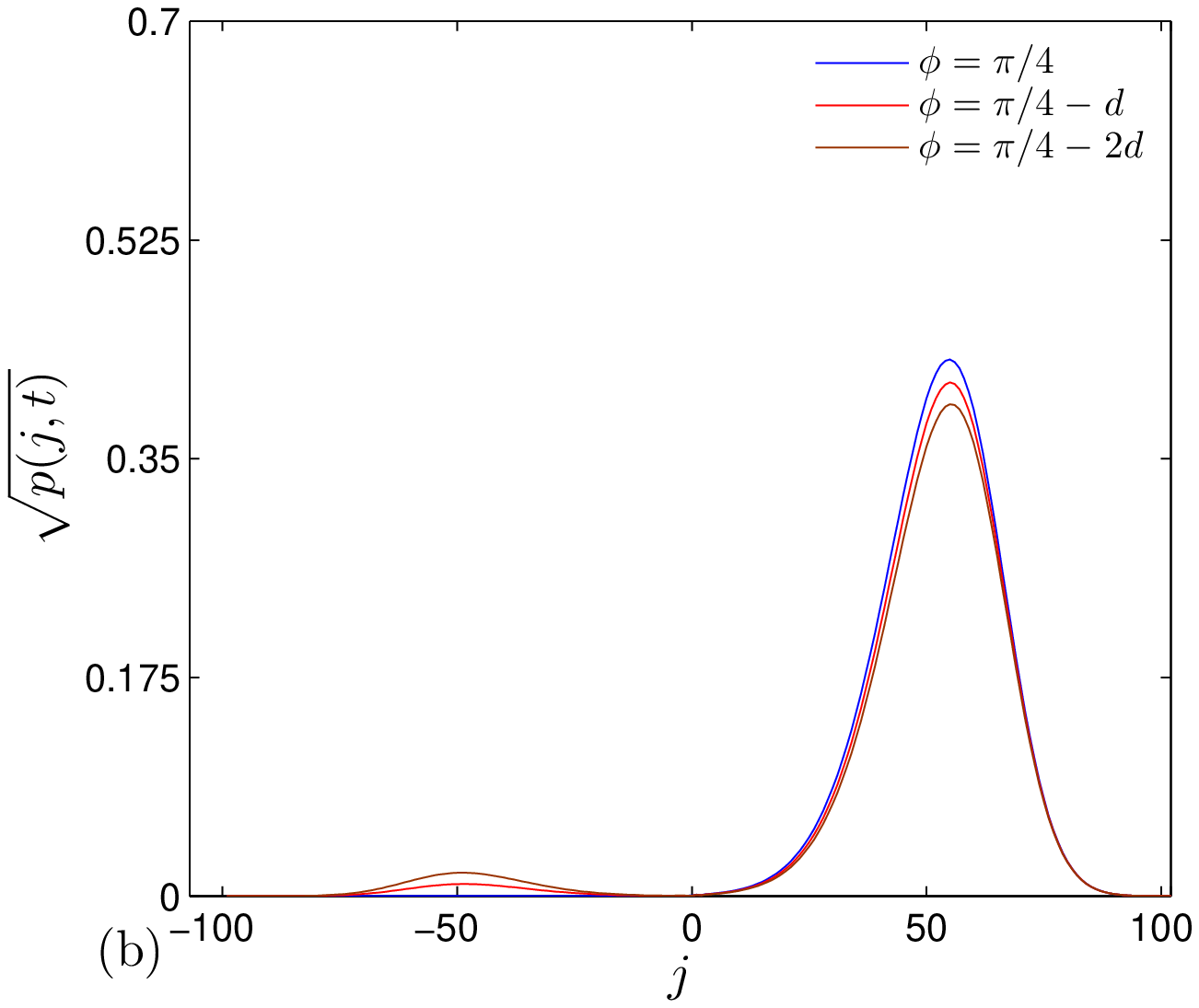} %
\includegraphics[ bb=39 238 420 558, width=0.4\textwidth, clip]{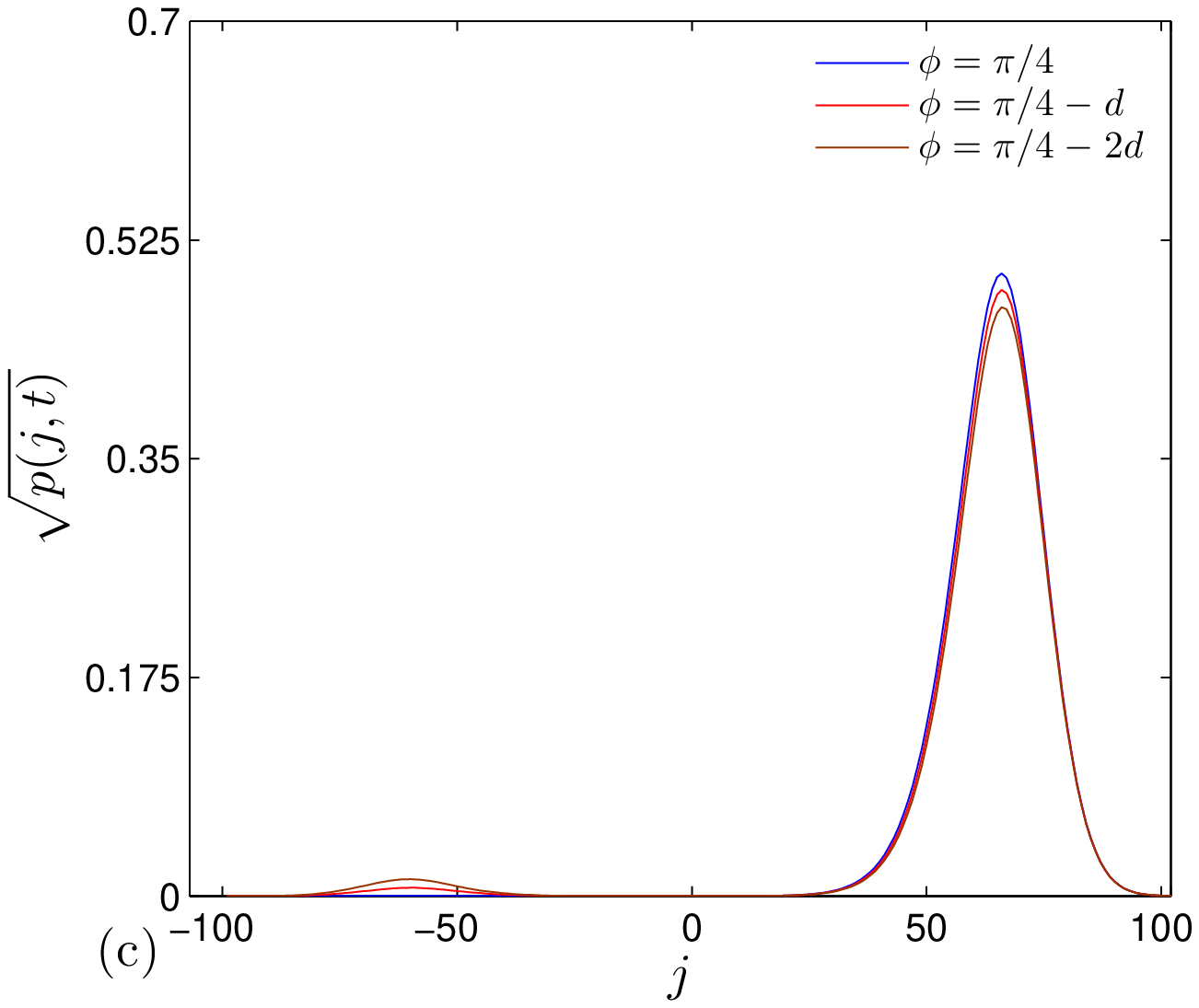} %
\includegraphics[ bb=39 238 420 558, width=0.4\textwidth, clip]{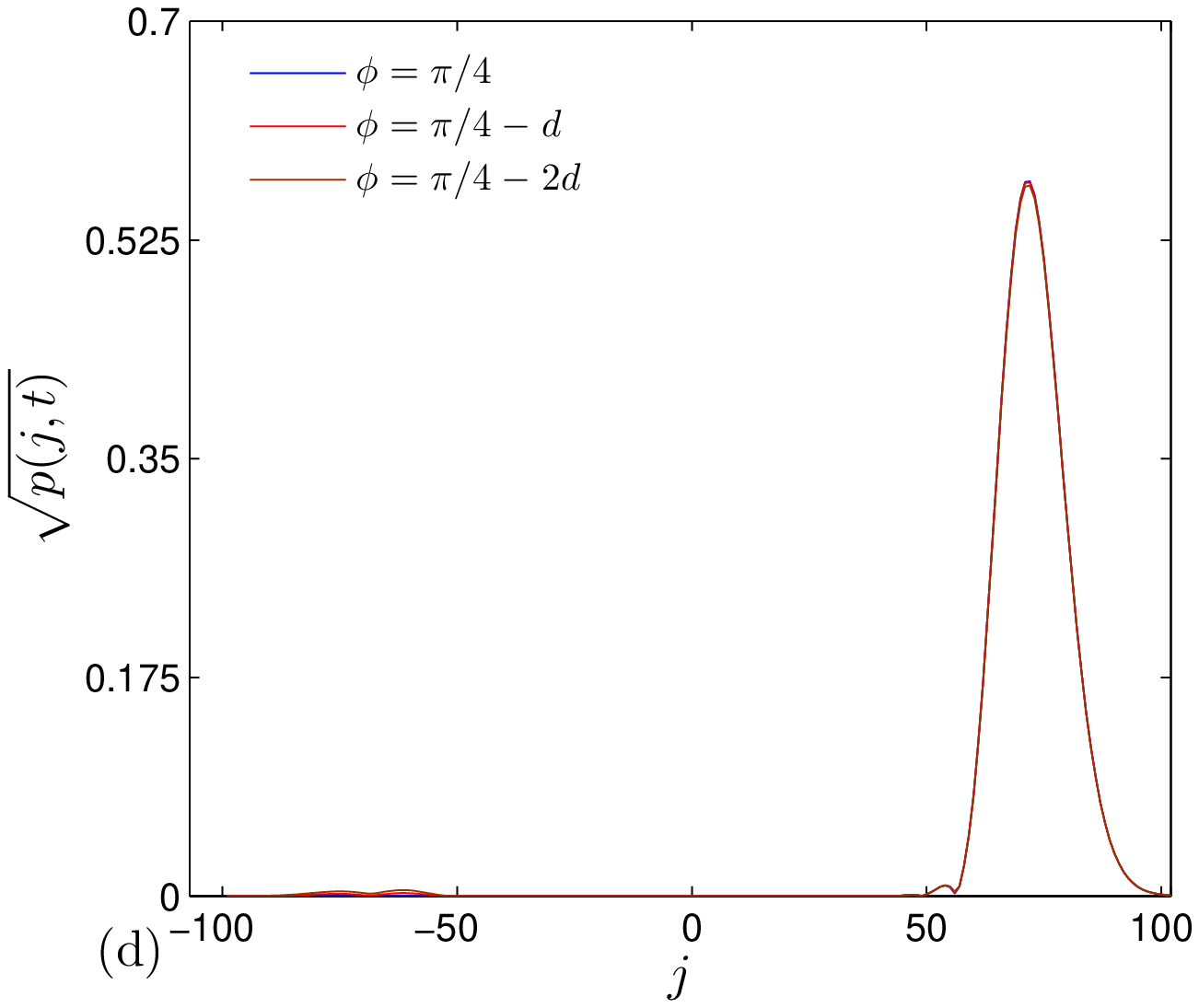}
\caption{(color online) The profiles of Dirac norm for time evolution of
initial Gaussian wave packets with $w=0.15$ and several typical $k_{0}$
under the systems with parameters $\protect\delta =-1.25$ and $\protect%
\gamma =0.75$ but different fluxes. (a) $k_{0}=\protect\pi /2$ and $\protect%
\phi =\protect\pi /4$. The blue line represents the time evolution at
several typical instants. As a comparison, the red circle shows the time
evolution of the same initial state under the system with parameters $%
\protect\delta =-1$ and $\protect\gamma =0$, which corresponds to a uniform
chain. It shows a perfect reflectionless amplification with $\protect\nu =2$%
. Figs. (b, c, d) show the time evolution at instant $t=70$ for different
initial states with (b) $k_{0}=\protect\pi /3$, (c) $k_{0}$ $=\protect\pi %
/2.5$, and (d) $k_{0}=\protect\pi /2$, respectively. Here $\protect\phi $
are deviated from $\protect\pi /4$\ in unit of $d=\protect\pi /100$. It
shows that this amplifier is immune of the deviation of $\protect\phi $\ for
the signal around $k_{0}=\protect\pi /2$.}
\label{fig3}
\end{figure*}

For an incident plane wave with momentum $k$ incoming from the left with
energy $E_{k}=-\left( e^{ik}+e^{-ik}\right) $, the scattering wave function $%
\left\vert \psi _{k}\right\rangle $ can be obtained by the Bethe Ansatz
method. The wave function has the form%
\begin{eqnarray}
\left\vert \psi _{k}\right\rangle &=&\sum_{j=1}^{\infty }[f^{k}\left(
j\right) \left\vert j\right\rangle +f^{k}\left( -j\right) \left\vert
-j\right\rangle ]  \notag \\
&&+f^{k}\left( \alpha \right) \left\vert \alpha \right\rangle +f^{k}\left(
\beta \right) \left\vert \beta \right\rangle
\end{eqnarray}%
where the scattering wave function $f^{k}\left( j\right) $ is in the form of
\begin{equation}
f^{k}\left( j\right) =\left\{
\begin{array}{cc}
e^{ikj}+r_{k}e^{-ikj}, & j\leqslant -1 \\
t_{k}e^{ik\left( j+1\right) }, & j\geq 1 \\
1+r_{k}, & j=\alpha \\
t_{k}e^{ik}, & j=\beta%
\end{array}%
\right. .
\end{equation}%
Here $r_{k}$ and $t_{k}$ are the reflection and transmission amplitudes of
the incident wave with momentum $k$, which can be used to identify the
spectral singularity of the system. By solving the Schr\"{o}dinger equation $%
H\left\vert \psi _{k}\right\rangle =E_{k}\left\vert \psi _{k}\right\rangle $%
, we obtain%
\begin{equation}
r_{k}=\frac{1-\mu \nu }{\mu \nu -e^{-i2k}}\text{, }t_{k}=\frac{\nu \left(
1-e^{-i2k}\right) }{\mu \nu -e^{-i2k}}\text{.}  \label{t_k}
\end{equation}%
Similarly, the solution for an incident wave from right can be obtained by
replacing $\nu $\ with\ $\mu $.

\begin{figure*}[tbp]
\centering
\includegraphics[ bb=53 238 420 554, width=0.4\textwidth, clip]{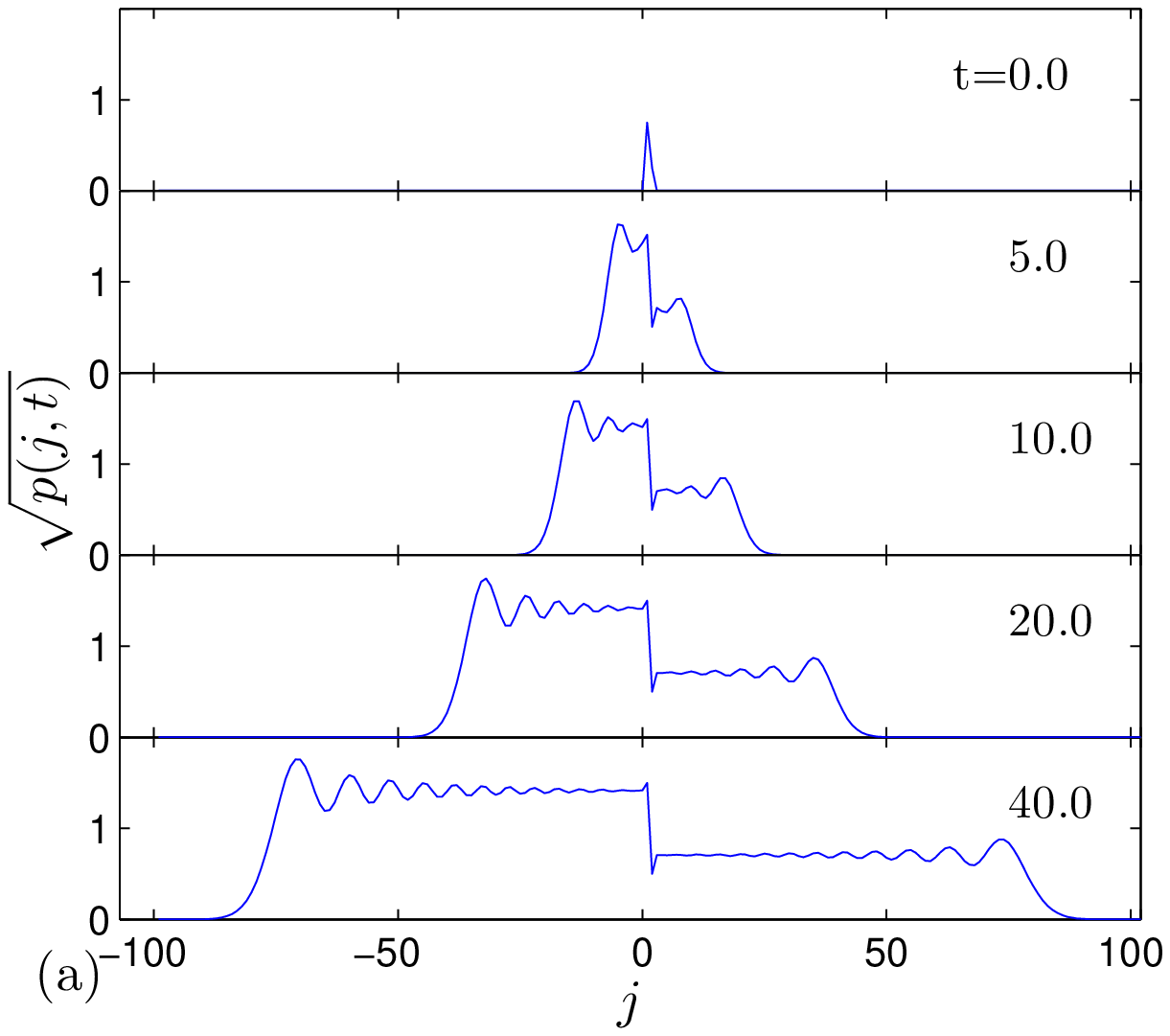} %
\includegraphics[ bb=53 238 420 554, width=0.4\textwidth, clip]{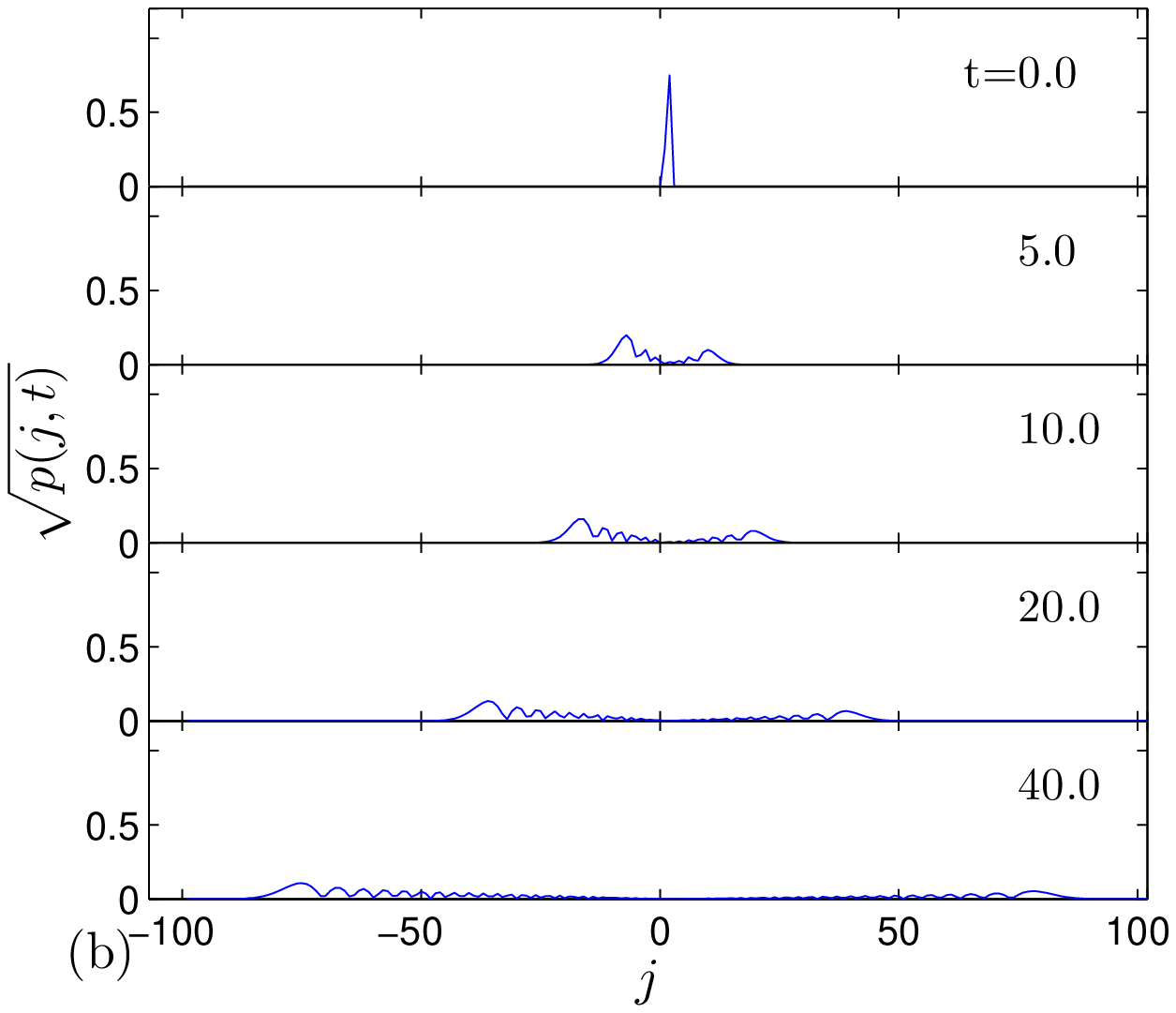} %
\includegraphics[ bb=53 238 420 554, width=0.4\textwidth, clip]{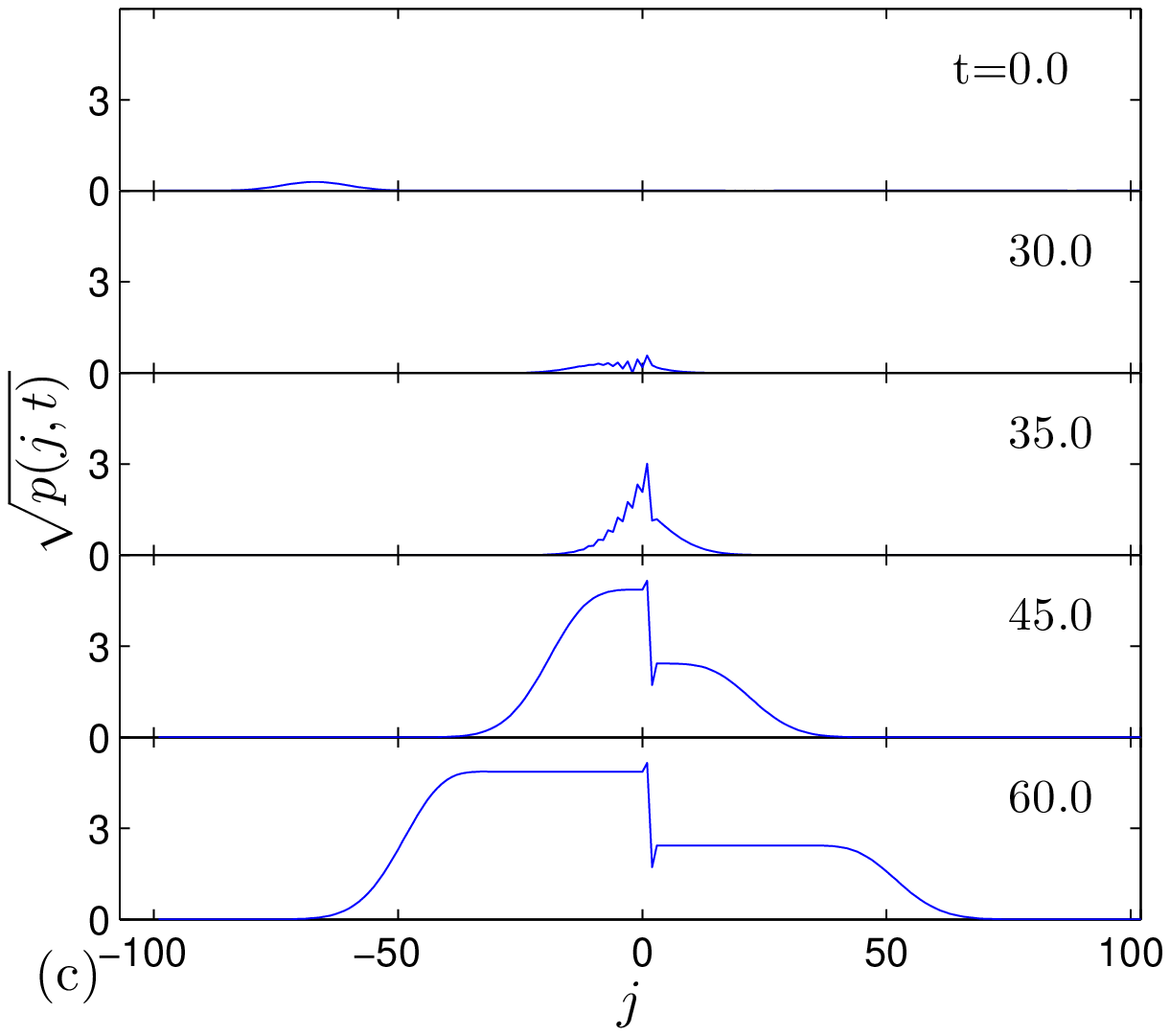} %
\includegraphics[ bb=53 238 420 554, width=0.4\textwidth, clip]{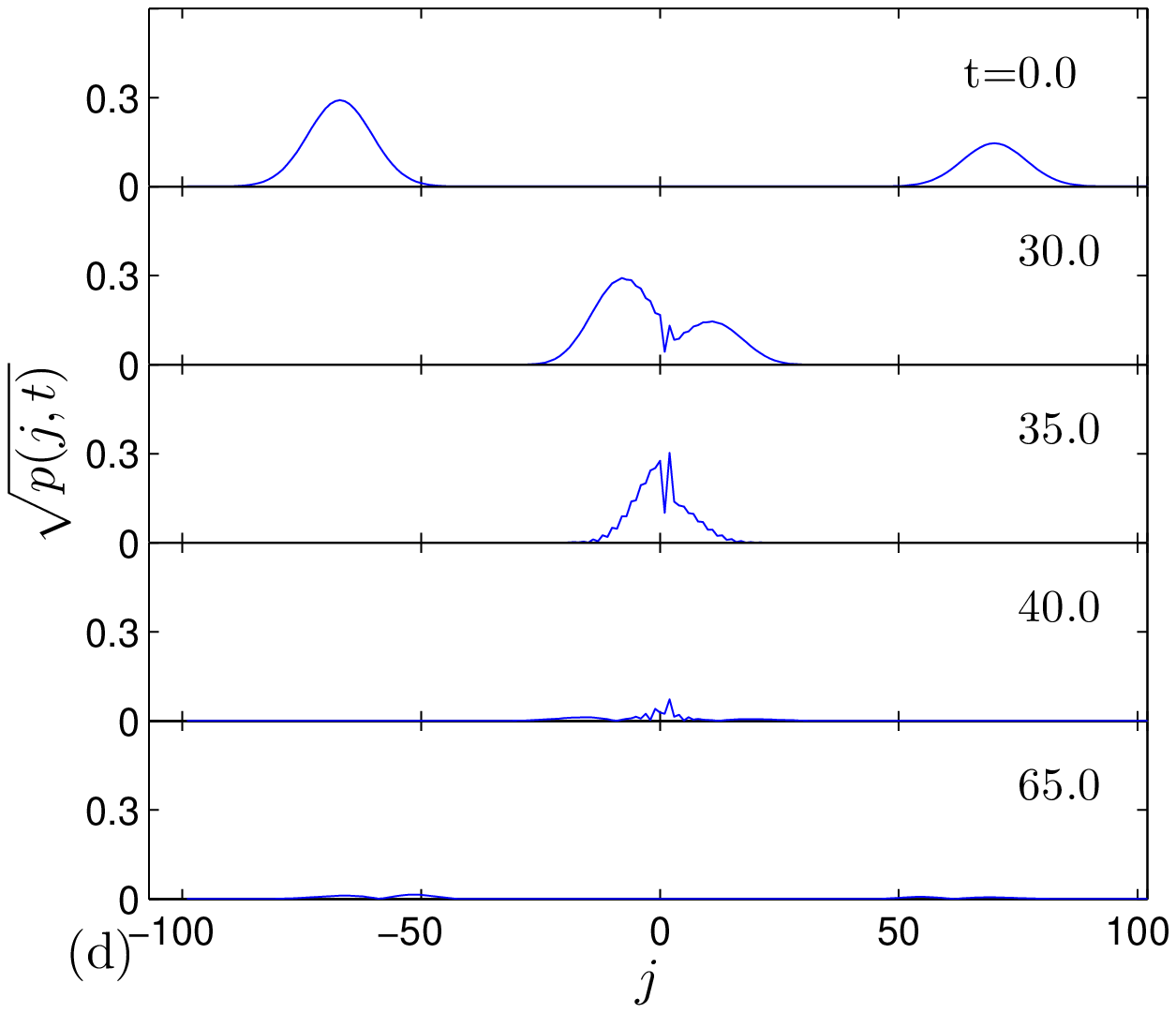}
\caption{(color online) The profiles of Dirac norm for evolved states of
four types of initial states: (a) The state defined by $\left\vert \protect%
\alpha \right\rangle +i\protect\nu \left\vert \protect\beta \right\rangle $;
(b) The state defined by $\left\vert \protect\alpha \right\rangle -i\protect%
\nu \left\vert \protect\beta \right\rangle $; (c) A Gaussian wave packet
with $w=0.15$ and $k_{0}=\protect\pi /2$; (d) The state defined by Eq. (%
\protect\ref{waves}) with $w=0.15$ and $k_{0}=\protect\pi /2$. The time
evolutions is governed by the Hamiltonian with $\protect\delta =0.75$ and $%
\protect\gamma =1.25$ for all figures, which correspond to $\protect\mu %
=-2.0 $ and $\protect\nu =0.5$, satisfying the spectral singularity
condition $\protect\mu \protect\nu =-1.0$.}
\label{fig4}
\end{figure*}

\section{Reflectionless amplification}

\label{Reflectionless amplification}

Besides the above property, Eq. (\ref{t_k}) also indicate the relation

\begin{equation}
r_{k}=0\text{, }t_{k}=\mu \text{,}
\end{equation}%
for left incident wave, or

\begin{equation}
r_{k}=0\text{, }t_{k}=\nu \text{,}
\end{equation}%
for right incident wave when the resonant condition $\mu \nu =1$ is
satisfied. The action of a resonant\ asymmetric dimer is reflectionless
amplified transmission. We note that this amplification is $k$-independent,
which results in deformation free for arbitrary signal.

This feature can be understood by the Eq. (\ref{heq_1}). We are interested
in the case of $\mu \nu =1$. One can rewrite the Hamiltonian $h_{\mathrm{eq}%
} $ as%
\begin{eqnarray}
h_{\mathrm{eq}} &=&-\sum_{j=1}^{\infty }(\overline{\left\vert j\right\rangle
}\overline{\left\langle j+1\right\vert }+\overline{\left\vert
j+1\right\rangle }\overline{\left\langle j\right\vert }+\overline{\left\vert
-j\right\rangle }\overline{\left\langle -j-1\right\vert }  \notag \\
&&+\overline{\left\vert -j-1\right\rangle }\overline{\left\langle
-j\right\vert })-(\overline{\left\vert -1\right\rangle }\overline{%
\left\langle \alpha \right\vert }+\overline{\left\vert \alpha \right\rangle }%
\overline{\left\langle -1\right\vert }  \notag \\
&&+\overline{\left\vert 1\right\rangle }\overline{\left\langle \beta
\right\vert }+\overline{\left\vert \beta \right\rangle }\overline{%
\left\langle 1\right\vert })-\overline{\left\vert \alpha \right\rangle }%
\overline{\left\langle \beta \right\vert }-\overline{\left\vert \beta
\right\rangle }\overline{\left\langle \alpha \right\vert },  \label{heq_2}
\end{eqnarray}%
which is equivalent to a Hermitian uniform chain. It accords with the fact
of $r_{k}=0$. When we concern about the Dirac probability, we can take the
mapping $\overline{\left\vert j\right\rangle }\rightarrow \sqrt{\frac{\nu }{%
\mu }}\left\vert j\right\rangle =\nu \left\vert j\right\rangle $, within the
right half region.

It indicates that the cluster in Eq. (\ref{original}) can act as a quantum
amplifier under the condition%
\begin{equation}
\mu \nu =\delta ^{2}-\gamma ^{2}=1.
\end{equation}%
We introduce the amplification coefficient $\mathcal{A}$\ to characterize
this phenomenon for an incident plane wave $\left\vert k,\text{in}%
\right\rangle $\ and output plane wave $\left\vert k,\text{out}\right\rangle
$, where
\begin{equation}
\mathcal{A}(k)=\frac{\langle k,\text{out}\left\vert k,\text{out}%
\right\rangle }{\langle k,\text{in}\left\vert k,\text{in}\right\rangle }.
\label{A}
\end{equation}%
A remarkable feature of this design is that
\begin{equation}
\mathcal{A}(k)=\nu ^{2}\text{, }\frac{\partial \mathcal{A}(k)}{\partial k}=0,
\label{k-indep}
\end{equation}%
i.e., the amplification coefficient is $k$-independent, which\ leads to the
fact that it cannot induce any distortion of a given signal. To demonstrate
this point, we perform numerical simulations for the scattering center in
Eq. (\ref{original}). The profiles of evolved states, i.e., $p\left(
j,t\right) =\left\vert \left\langle j\right\vert \varphi \left( t\right)
\rangle \right\vert ^{2}$, which is the probability of the evolved state $%
\left\vert \varphi \left( t\right) \right\rangle $ on the site $j$, are
plotted in Fig. \ref{fig3}(a), which shows that the wave packet $\left\vert
\varphi \left( 0\right) \right\rangle =\left\vert \phi \left( N_{A},\pi
/2\right) \right\rangle $ almost totally passes through the scattering
center and the transmitted wave packet is amplified by $\nu $ times of the
incident amplitude without any signal distortion. Where the initial state $%
\left\vert \phi \left( N_{A},k_{0}\right) \right\rangle =\Omega
_{0}^{-1/2}\sum_{j}e^{-\frac{\lambda ^{2}}{2}\left( j-N_{A}\right)
^{2}}e^{ik_{0}j}\left\vert j\right\rangle $ represents a Gaussian wave
packet with central momentum $k_{0}$\ and position $N_{A}$, $\Omega
_{0}=\sum_{j}e^{-\lambda ^{2}\left( j-N_{A}\right) }$\ is the normalization
factor, and the half-width of the wave packet $w=2\sqrt{\ln 2}/\lambda $
characterizes the size of the local state.

For the experimental realization of quantum amplifier of this design, the
deviation of a magnetic flux from the optimal magnitude may change the Eq. (%
\ref{k-indep}) in practice. Figs. \ref{fig3}(b,c,d) is the plots of the
profiles of evolved Gaussian wave packets under the system with the
deviated\ flux, which are obtained by numerical simulations for Gaussian
wave packets with several typical values of central momenta $k_{0}$. It
indicates that the deviation of the flux results in the distortion of the
shape. One can find that the more $\phi $\ and $k_{0}$ deviate from $\pi /4$
and $\pi /2$, respectively, the more distorted signal is obtained from the
amplifier. It shows that the quantum amplifier of this design is robust with
respect to the deviation of flux.

\begin{figure}[tbp]
\includegraphics[bb=30 444 550 790, width=0.43\textwidth, clip]{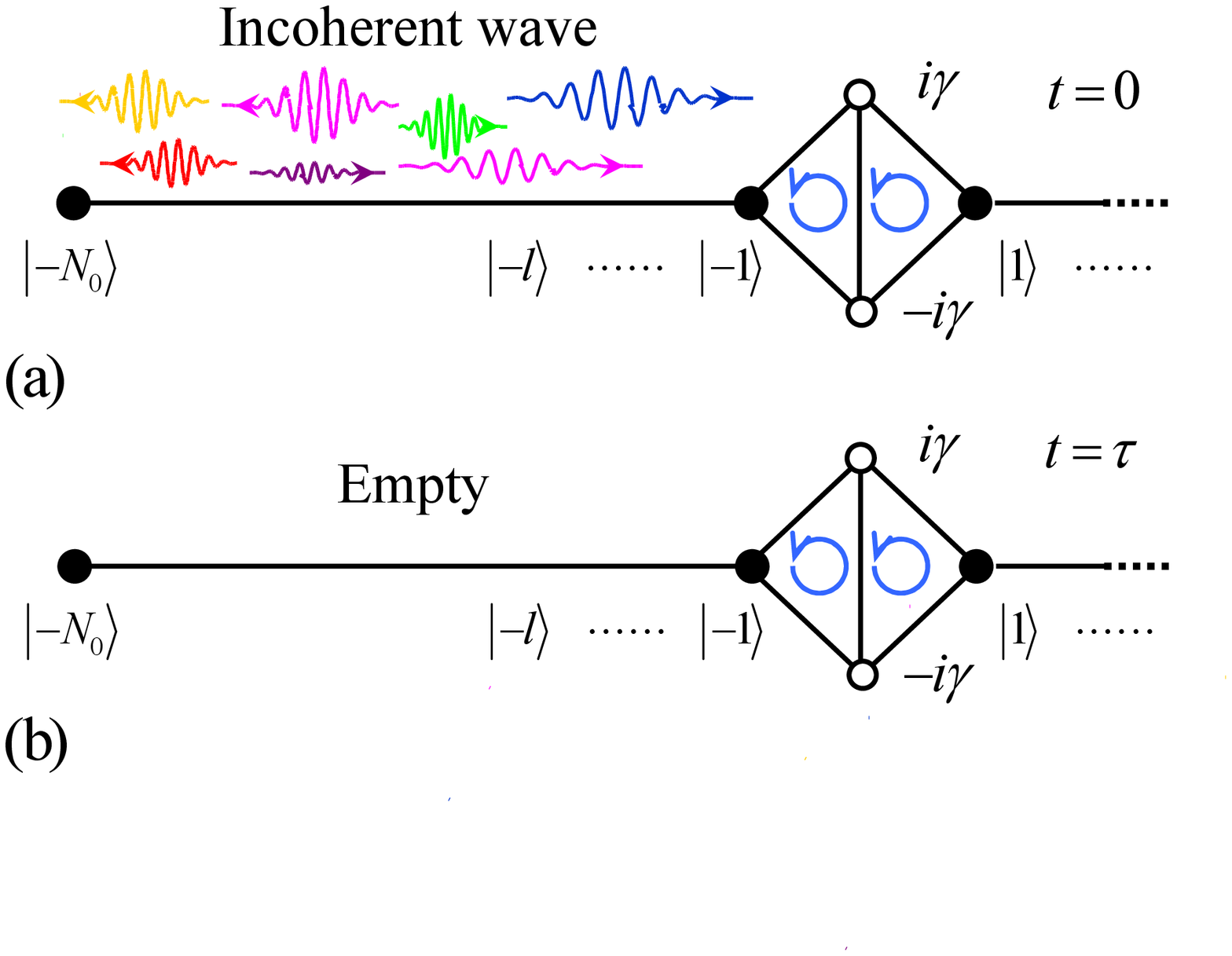}
\caption{(Color online) Schematic illustration of a non-Hermitian
configuration to demonstrate the perfect absorption for incoherent state. It
is the same system as Fig. \protect\ref{figure2}(a) with open boundary
condition at left side. (a) Initially, a mixed state is located in the
region $[-N_{0},-1]$. (b) The incoherent perfect absorption is achieved if
the total Dirac probability in the whole system vanishes at relaxation time $\protect\tau $.}
\label{fig5}
\end{figure}

\begin{figure}[tbp]
\includegraphics[bb=95 239 442 510, width=0.43\textwidth, clip]{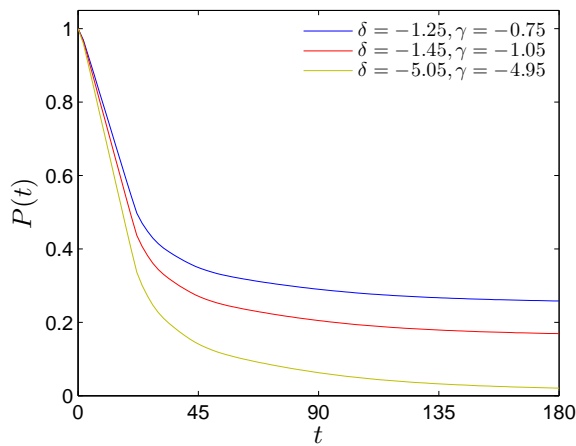}
\caption{(Color online) The plots of $P(t)$ for the initial state in Eq. (\protect\ref{initial state}), obtained from the time evolution under the
configuration illustrated in Fig. \protect\ref{fig5} with $N_{0}=20$. The
inset parameters $(\protect\delta ,\protect\gamma )$ corresponds to $\protect\nu =0.5$, $0.4$, and $0.1$, respectively.}
\label{fig6}
\end{figure}

\section{Spectral singularity}

\label{Spectral singularity}

Recently, the concept of spectral singularity of a non-Hermitian system has
gained a lot of attention \cite%
{PRA1,PRB1,Ali3,PRA3,JMP5,PRD1,PRA4,PRA5,PRA6,Fagotti,Kun}, motivated by the
possible physical relevance of this since the pioneer work of\ Mostafazadeh
\cite{PRL3}.\ The majority of previous works do not include the
non-Hermitian system with asymmetric dimer. According to Eq. (\ref{t_k}),
The divergence of $r_{k}$ and $t_{k}$\ indicates the existence of spectral
singularity, which has been pointed out in Ref. \cite{ZXZ PRA2013}. We note
that it occurs only at the point $\mu \nu =-1$ for states with $k=\pm \pi /2$%
. The corresponding wave functions are%
\begin{equation}
f^{\pm \pi /2}\left( j\right) =\left\{
\begin{array}{cc}
e^{i\left( \pm \pi /2\right) j}, & j\leqslant -1 \\
\nu e^{i\left( \mp \pi /2\right) \left( j+1\right) }, & j\geq 1 \\
1, & j=\alpha \\
\nu e^{i\left( \mp \pi /2\right) }, & j=\beta%
\end{array}%
\right. .  \label{SS WF}
\end{equation}%
The physics of two states are clear, representing complete absorption and
self-sustained emission of two opposite waves. The intriguing feature of the
spectral singularity is the coexistence of two states. It indicates that the
cluster in Eq. (\ref{original}) can be a source and drain simultaneously
when we take%
\begin{equation}
\mu \nu =\delta ^{2}-\gamma ^{2}=-1.
\end{equation}%
We note that $\left\vert \gamma \right\vert >1$\ is the necessary condition
for the existence of the spectral singularity. For the simplest case with $%
\left\vert \gamma \right\vert =1$ and $\delta =0$, the system reduces to the
one which has been systematically studied in Ref. \cite{ZG PRA}.\

To demonstrate this point, we perform numerical simulations for the
scattering center in Eq. (\ref{original})\ with several typical situations.
First of all, the wave functions in Eq. (\ref{SS WF}) show that the state $%
\left\vert \alpha \right\rangle +i\nu \left\vert \beta \right\rangle $
should trigger a self-sustained emission of two counter propagating waves;
while the state $\left\vert \alpha \right\rangle -i\nu \left\vert \beta
\right\rangle $ should damp\ as time evolution since it cannot stimulate
waves from both the left and the right of infinity. Obviously, any linear
superposition of above two states still can be a seed state to generate two
waves prorogating to the left and the right of infinity. Similarly, any pair
of state $i^{j}\left\vert -j\right\rangle +\nu i^{\left( j+1\right)
}\left\vert j\right\rangle $ and $i^{-j}\left\vert -j\right\rangle +\nu
i^{-\left( j+1\right) }\left\vert j\right\rangle $ ($j>0$)\ have the same
dynamical property, i.e., the former can be a seed of persistent waves;
while the later should be absorbed partially by the center. In Fig. \ref%
{fig4}(a,b), we stimulate the time evolution $\left\vert \varphi \left(
t\right) \right\rangle =e^{-iHt}\left\vert \varphi \left( 0\right)
\right\rangle $ of the initial states$\ $%
\begin{equation}
\left\vert \varphi \left( 0\right) \right\rangle =\left\vert \alpha
\right\rangle +i\nu \left\vert \beta \right\rangle  \label{position1}
\end{equation}%
and%
\begin{equation}
\left\vert \varphi \left( 0\right) \right\rangle =\left\vert \alpha
\right\rangle -i\nu \left\vert \beta \right\rangle .  \label{position2}
\end{equation}%
We see that the wave packet dynamics exhibits distinct spectral-singularity
characteristics. In Fig. \ref{fig4}(a), state $\left\vert \alpha
\right\rangle +i\nu \left\vert \beta \right\rangle $ stimulates two counter
propagating waves with the amplitude ratio for the left side to right side
as $1:\nu $; while (b) state $\left\vert \alpha \right\rangle -i\nu
\left\vert \beta \right\rangle $ spreads out as a short-lived seed state. It
indicates that these two local states (non-orthogonal states in the
framework of Dirac inner product) show total different behavior, which may
be exploited to the scheme for detecting a specific state.

Secondly, any asymptotic plane-wave solution can be demonstrated by the
dynamics of wavepacket. There are two typical solutions: (i) Consider a
general single-incident solution, which has the form
\begin{equation}
(e^{-ikj}+r_{k}e^{ikj})\left\vert -j\right\rangle +t_{k}e^{ik\left(
j+1\right) }\left\vert j\right\rangle
\end{equation}%
with $1<j\leqslant N$.\ The reflection and transmission go to infinity for $%
k=\pi /2$. This solution corresponds to the wave emission dynamics of the
initial state%
\begin{equation}
\left\vert \varphi \left( 0\right) \right\rangle =\left\vert \phi \left(
N_{A},\pi /2\right) \right\rangle ,  \label{Gauss}
\end{equation}%
\ The infinity of $r_{k}$\ and $t_{k}$\ should exhibit in the dynamics of
the wave packet.\textbf{\ }(ii) The solution $f^{\pi /2}\left( j\right) $
corresponds to the dynamics of two counter propagating wave packets
initially centered at $\pm N_{A}$%
\begin{equation}
\left\vert \varphi \left( 0\right) \right\rangle =\left\vert \phi \left(
-N_{A},k_{0}\right) \right\rangle -i\nu \left\vert \phi \left(
N_{A},-k_{0}\right) \right\rangle ,  \label{waves}
\end{equation}%
which is anti-symmetric with respect to the scattering center. The profiles
of the Dirac norm of evolved states $\left\vert \varphi \left( t\right)
\right\rangle $, i.e., $p\left( j,t\right) =\left\vert \left\langle
j\right\vert \varphi \left( t\right) \rangle \right\vert ^{2}$, which is the
probability of the state $\left\vert \varphi \left( t\right) \right\rangle $
on the site $j$, computed by exact numerical diagonalization, are plotted in
Figs. \ref{fig4}(c,d).

In Fig. \ref{fig4}(c) an incident wave packet stimulates two counter
propagating emission waves with the amplitude ratio $1:\nu $. The Dirac
probabilities of reflected and transmitted waves increase linearly as time
goes on, which is a dynamical demonstration of the infinite reflection and
transmission coefficients. In Fig. \ref{fig4}(d), a typical example of
complete absorption is shown. Two incident waves with matching amplitudes
and relative phases are fully absorbed after scattering.

In fact, all the simulation results can be understood by a equivalent model.
We introduce the transformation%
\begin{equation}
\overline{\left\vert j\right\rangle }=\left\{
\begin{array}{cc}
\sqrt{\nu /\mu }\left\vert j\right\rangle , & \left( j\geqslant 1,j=\beta
\right) \\
\left\vert j\right\rangle , & \left( j\leqslant 1,j=\alpha \right)%
\end{array}%
\right. ,
\end{equation}%
and its conjugation%
\begin{equation}
\overline{\left\langle j\right\vert }=\left\{
\begin{array}{cc}
\sqrt{\mu /\nu }\left\langle j\right\vert , & \left( j\geqslant 1,j=\beta
\right) \\
\left\langle j\right\vert , & \left( j\leqslant 1,j=\alpha \right)%
\end{array}%
\right. ,
\end{equation}%
which satisfy the biorthonormal relation%
\begin{equation}
\overline{\left\langle j\right\vert l\rangle }=\delta _{jl}.
\end{equation}%
One can rewrite the Hamiltonian $H_{\mathrm{eq}}$\ as%
\begin{eqnarray}
h_{\mathrm{eq}} &=&-\sum_{j=1}^{\infty }\left( \overline{\left\vert
j\right\rangle }\overline{\left\langle j+1\right\vert }+\overline{\left\vert
-j\right\rangle }\overline{\left\langle -j-1\right\vert }\right)  \notag \\
&&-\left( \overline{\left\vert -1\right\rangle }\overline{\left\langle
\alpha \right\vert }+\overline{\left\vert 1\right\rangle }\overline{%
\left\langle \beta \right\vert }\right) +\mathrm{H.c}.  \notag \\
&&-\sqrt{\mu \nu }\overline{\left\vert \alpha \right\rangle }\overline{%
\left\langle \beta \right\vert }-\sqrt{\mu \nu }\overline{\left\vert \beta
\right\rangle }\overline{\left\langle \alpha \right\vert },  \label{heq_1}
\end{eqnarray}%
which becomes Hermitian in the case of $\mu \nu >0$. We would like to point
out that the above transformation is suitable for all range of parameters\ $%
(\mu ,\nu )$\ \cite{ZXZ Ann}.\textbf{\ }In the case of $\mu \nu =-1$,
although it is non-Hermitian, it has $\mathcal{P}$ symmetry, which allows us
to block-diagonalize the matrix.\ It is shown that (see Appendix b) $h_{%
\mathrm{eq}}$\ can be decomposed to two independent sub-Hamiltonians, which
represent semi-infinite chains with ending imaginary potential $i$ and $-i$%
,\ respectively. According to Refs. \cite{Longhi1,ZXZ PRA2013}, each
sub-Hamiltonian has its own spectral singularity. It provides a clear
physical picture for understanding the coexistence of two spectral
singularities in the original system $H$ at the point $\gamma ^{2}-\delta
^{2}=1$.

\section{Absorption of incoherent wave}

\label{Absorption of incoherent wave} We would like to address that the $k$%
-independent reflectionless transmission is unconditional. It is based on
the fact that the equivalent Hamiltonian Eq. (\ref{heq_2}) is\ applicable
for an arbitrary state, including a mixed state. On the other hand, in the
limit case $\nu \rightarrow 0$, or $\gamma -\delta \ll \delta ^{2}-\gamma
^{2}=1$, the transmitted probability is attenuated to nothing. In this
sense, the non-Hermitian scattering center acts as a perfect absorber. Note
that by combining two above features, we find that it presents an incoherent
perfect absorption.

In order to clearly demonstrate this point, we consider the system with the
modified lead Hamiltonian

\begin{equation}
H_{\mathrm{lead}}=-\sum_{j=1}^{\infty }\left\vert j\right\rangle
\left\langle j+1\right\vert -\sum_{j=N_{0}}^{\infty }\left\vert
-j\right\rangle \left\langle -j-1\right\vert +\mathrm{H.c.}.
\end{equation}%
We take an open boundary condition at $N_{0}$-site, in order to avoid the
particle probability escaping to the left. We examine the function of the
scattering center by calculating\ the time evolution of a given mixed state
located in the region $[-N_{0},-1]$. Intuitively, any local state near the
one end of the semi-infinite uniform chain should spread out to the right
infinitely. In principle, it is due to the absence of bound state in a
semi-infinite uniform chain.

In general, a mixed state is described by a density matrix $\rho \left(
t\right) $, which obeys the Schr\"{o}dinger equation%
\begin{equation}
i\frac{\partial \rho \left( t\right) }{\partial t}=\left[ H,\rho \left(
t\right) \right] .
\end{equation}%
The solution of the equation has the form%
\begin{equation}
\rho \left( t\right) =e^{-iHt}\rho \left( 0\right) e^{iH^{\dagger }t},
\end{equation}%
which is basis\ for numerical simulation in the following. The Dirac
probability at $j$-th site can be obtained as%
\begin{equation}
p\left( j,t\right) =\mathrm{Tr}[\left\vert j\right\rangle \left\langle
j\right\vert \rho \left( t\right) ],
\end{equation}%
where $\mathrm{Tr}[...]$\ denotes the trace of a matrix. Then the total
probability at time $t$ is%
\begin{equation}
P(t)=\sum_{j=-N_{0}}^{\infty }p\left( j,t\right) .
\end{equation}%
We consider the time evolution of an initial mixed state density matrix
\begin{equation}
\rho \left( 0\right) =\frac{1}{N_{0}}\sum_{j=1}^{N_{0}}\left\vert
-j\right\rangle \left\langle -j\right\vert ,  \label{initial state}
\end{equation}%
under several typical parameters $\left\{ \gamma ,\delta \right\} $
satisfying $\delta ^{2}-\gamma ^{2}=1$. Fig. \ref{fig6} presents the plots
of the numerical results. It indicates that the probability in the whole
system drops rapidly within a certain period of time, and optimal parameters
can lead to near perfect absorption. State $\rho \left( 0\right) $\ contains
components that cover all possible $k$. Although we cannot give a proof, our
calculation indicates that such an absorber can be used to treat\ all kind
of states.

\section{Summary}

\label{Summary}

In summary, we have studied the non-Hermitian Aharonov--Bohm interferometer.
We have shown that the combination of imaginary potentials and magnetic flux
can result in asymmetric transmission, which is not achievable in principle
when one of them is solely present, which is not achievable when either one
is solely present. Inspired by this, we construct an equivalent
non-Hermitian dimer with an unequal hopping rate, which as a fundamental
non-Hermitian element has been shown to have many intriguing features,\ by
an interferometer-like\ cluster in the framework of tight-binding model. It
is the first time to establish an exact equivalence between two
non-Hermitian building blocks, which paves the way for the non-Hermitian
device design. As an application, this configuration can act as a
unidirectional quantum amplifier. The remarkable\ feature of this design is
wave-vector independent, which allows the reflectionless amplified
transmission of a signal without any distortion. Furthermore, with optimal
system parameters, it acts as an absorber for both coherent and incoherent
incident waves, which may be applicable to photovoltaic or stealth
technology. In addition, we investigate the dynamical behaviors related to
the spectral singularities analytically and numerically.

\appendix*

\section{a. On-site potential scattering}

We consider a general scattering center, an on-site potential $V$\ which may
be real or complex, embedded in the center of an infinite chain with the
Hamiltonian%
\begin{equation}
H_{\mathrm{V}}=H_{\mathrm{lead}}+H_{\mathrm{c}}
\end{equation}%
where%
\begin{equation}
H_{\mathrm{lead}}=-\sum_{j=1}^{\infty }\left( \left\vert j\right\rangle
\left\langle j+1\right\vert +\left\vert -j\right\rangle \left\langle
-j-1\right\vert +\mathrm{H.c.}\right) ,
\end{equation}%
is the Hamiltonian of two leads and the scattering center Hamiltonian%
\begin{equation}
H_{\mathrm{c}}=-\left( \left\vert -1\right\rangle +\left\vert 1\right\rangle
\right) \left\langle 0\right\vert +\mathrm{H.c.}+V\left\vert 0\right\rangle
\left\langle 0\right\vert .
\end{equation}%
The Bethe Ansatz solution has the form%
\begin{equation}
(e^{-ikj}+r_{k}e^{ikj})\left\vert -j\right\rangle +t_{k}e^{ik\left(
j+1\right) }\left\vert j\right\rangle ,\text{ }(j>0)
\end{equation}
By solving the Schr\"{o}dinger equation \cite{Kim PRB}, the transmission and
reflection amplitudes are obtained as

\begin{equation}
t_{k}=\frac{2i\sin k}{2i\sin k-V},\text{ }r_{k}=\frac{V}{2i\sin k-V}.
\end{equation}%
For real $V$, the transmission and reflection coefficients $T_{k}=\left\vert
t_{k}\right\vert ^{2}$, $R_{k}=\left\vert r_{k}\right\vert ^{2}$\ are
expressed as

\begin{equation}
T_{k}=\frac{4\sin ^{2}k}{4\sin ^{2}k+V^{2}},\text{ }R_{k}=\frac{V^{2}}{4\sin
^{2}k+V^{2}},
\end{equation}%
while an imaginary potential $V=i\gamma $ leads to%
\begin{equation}
T_{k}=\frac{4\sin ^{2}k}{\left( 2\sin k-\gamma \right) ^{2}},\text{ }R_{k}=%
\frac{\gamma ^{2}}{\left( 2\sin k-\gamma \right) ^{2}}.
\end{equation}%
\bigskip

\section{b. Double spectral singularities}

The Hamiltonian $H$\ has double spectral singularities, which can be
understood from following derivation. Introducing a linear transformation%
\begin{eqnarray}
\overline{\left\vert -l\right\rangle } &=&\frac{1}{\sqrt{2}}\left( \overline{%
\left\vert j\right\rangle }+\overline{\left\vert -j\right\rangle }\right) ,%
\underline{\left\vert -l\right\rangle }=\frac{1}{\sqrt{2}}\left( \overline{%
\left\vert j\right\rangle }-\overline{\left\vert -j\right\rangle }\right) ,
\notag \\
\overline{\left\vert 0\right\rangle } &=&\frac{1}{\sqrt{2}}\left( \overline{%
\left\vert \alpha \right\rangle }+\overline{\left\vert \beta \right\rangle }%
\right) ,\underline{\left\vert 0\right\rangle }=\frac{1}{\sqrt{2}}\left(
\overline{\left\vert \alpha \right\rangle }-\overline{\left\vert \beta
\right\rangle }\right) ,
\end{eqnarray}%
we have%
\begin{equation}
h_{\mathrm{eq}}=h_{\mathrm{+}}+h_{\mathrm{-}},
\end{equation}%
where

\begin{equation}
h_{\mathrm{+}}=-\sum_{l=1}^{\infty }(\overline{\left\vert -l\right\rangle }%
\overline{\left\langle -l-1\right\vert }+\mathrm{H.c}.)+i\overline{%
\left\vert 0\right\rangle }\overline{\left\langle 0\right\vert },
\end{equation}%
and%
\begin{equation}
h_{\mathrm{-}}=-\sum_{l=1}^{\infty }(\underline{\left\vert -l\right\rangle }%
\underline{\left\langle -l-1\right\vert }+\mathrm{H.c}.)-i\underline{%
\left\vert 0\right\rangle }\underline{\left\langle 0\right\vert }.
\end{equation}%
It is easy to check that%
\begin{equation}
\left[ h_{\mathrm{+}},h_{\mathrm{-}}\right] =0.
\end{equation}

The physical picture is clear that $h_{\mathrm{eq}}$\ can be decomposed to
two independent sub-Hamiltonian $h_{\mathrm{\pm }}$, which\ represents
semi-infinite chain with an ending imaginary potential $\pm i$. Each
sub-Hamiltonian has its own spectral singularity. It is essential for the
double spectral singularities in the original system $H$ at the point $%
\gamma ^{2}-\delta ^{2}=1$.

Accordingly, the dynamical behavior of spectral singularities for $h_{%
\mathrm{\pm }}$\ can be demonstrated by the time evolutions of following
typical initial states. (i) Position state $\overline{\left\vert
0\right\rangle }$, which corresponds to state in Eq. (\ref{position1}). (ii)
Position state $\underline{\left\vert 0\right\rangle }$, which corresponds
to state in Eq. (\ref{position2}). (ii) Gaussian wave packet $\Omega
_{0}^{-1/2}\sum_{j<-1}$ $e^{-\frac{\lambda ^{2}}{2}\left( j-N_{A}\right)
^{2}}e^{ik_{0}j}$\underline{$\left\vert j\right\rangle $}, which corresponds
to state in Eq. (\ref{Gauss}).

\acknowledgments We acknowledge the support of the National Basic Research
Program (973 Program) of China under Grant No. 2012CB921900, CNSF (Grant No.
11374163), NSFC (Grant No. 11605094) and the Baiqing Plan of Nankai
University.

\end{document}